\documentclass[5p]{elsarticle} 

\usepackage{mathtools}
\journal{ }

\usepackage{placeins}

\usepackage{eurosym}
\usepackage{threeparttable} 

\usepackage[final]{pdfpages}

\usepackage{url}
\usepackage[colorlinks=true, citecolor=blue, linkcolor=blue, filecolor=blue,urlcolor=blue]{hyperref}

\usepackage{lineno}
\modulolinenumbers[5]

\usepackage{lineno,hyperref}
\modulolinenumbers[1]
\usepackage{amsmath}
\usepackage{siunitx}
\usepackage{eurosym}
\biboptions{numbers,sort&compress}
\usepackage[europeanresistors,americaninductors]{circuitikz}
\usepackage{adjustbox}
\usepackage{xspace}
\usepackage{caption}
\usepackage{booktabs}
\usepackage{tabularx}
\usepackage{multicol}
\usepackage{float}
\usepackage{graphicx,dblfloatfix}
\usepackage{csvsimple}

\begin{document}

\begin{frontmatter}

\title{Speed of technological transformations required in Europe to achieve different climate goals}
\author[mymainaddress,iClimate]{Marta Victoria\corref{mycorrespondingauthor}}
\ead{mvp@mpe.au.dk}
\author[kitaddress,TUaddress]{Elisabeth Zeyen}
\author[TUaddress]{Tom Brown}
\cortext[mycorrespondingauthor]{Corresponding author}
\address[mymainaddress]{Department of Mechanical and Production Engineering, Aarhus University, Inge Lehmanns Gade 10, 8000 Aarhus, Denmark}
\address[iClimate]{iCLIMATE Interdisciplinary Centre for Climate Change, Aarhus University}
\address[kitaddress]{Institute for Automation and Applied Informatics (IAI), Karlsruhe Institute of Technology (KIT), Forschungszentrum 449, 76344, Eggenstein-Leopoldshafen, Germany}
\address[TUaddress]{Department of Digital Transformation in Energy Systems, Technische Universit{\"a}t Berlin, Einsteinufer 25 (TA 8), 10587 Berlin, Germany}

\begin{abstract}
Europe's contribution to global warming will be determined by the cumulative emissions until climate neutrality is achieved. In this paper, we investigate alternative transition paths under carbon budgets corresponding to temperature increases between 1.5 and 2$^{\circ}$C. We use PyPSA-Eur-Sec, an open model of the sector-coupled European energy system with high spatial and temporal resolution. All the paths entail similar technological transformations, but the timing of the scale-up of important technologies like water electrolysis, carbon capture and hydrogen networks differs in the model. In our results, solar PV, onshore and offshore wind become the cornerstone of a net-zero energy system enabling the decarbonisation of other sectors via direct electrification (e.g. heat pumps and electric vehicles) or indirect electrification (e.g. using synthetic fuels). Under the cost and performance assumptions applied, for a social cost of carbon (SCC) of 120\EUR/tCO$_2$, transition paths under 1.5 and 1.6 $^{\circ}$C budgets are, respectively, 8\%, and 1\% more expensive than the 2$^{\circ}$C-budget because building assets earlier costs more. These pathways also see a faster ramp-up of new technologies before 2035. Under these assumptions, the 1.5$^{\circ}$C-budget is cost-optimal in our model, if SCC of at least 300 Euros are considered. Moreover, we discuss the strong implications of the SCC and discount rate assumed when comparing alternative paths. We also analyse the consequences of different assumptions on the cost and potential of CO$_2$ sequestration. 
\end{abstract}


\end{frontmatter}


\section{Introduction}

Sustained high global annual CO$_2$ emissions are quickly depleting our carbon budget, \textit{i.e.}, the cumulative emissions that will enable us to remain below a specific temperature increase. On top of that, estimating the carbon budget is subject to substantial uncertainties in the evaluation of the transient climate response to cumulative emissions or the potential impacts of Earth system feedbacks such as permafrost thawing \cite{Rogelj_2019}. An environmentally cautious approach would entail reducing emissions as fast as possible. In Europe, climate ambition has risen in recent years with examples such as the 55\% greenhouse gas (GHG) reduction commitment of the European Union for 2030 \cite{55target}, the European Green Deal \cite{GreenDeal}, and aggressive reduction targets in some member states \cite{Denmark2030, Germany2045}. Still, a large gap exists between globally committed emissions reductions and those necessary to fulfil the Paris Agreement \cite{Emissions_Gap}.

\

On the one hand, Integrated Assessment Models (IAMs) have traditionally been used to assess transition paths under strict carbon budgets \cite{Rogelj_2018, Vanvuuren_2018, Grubler_2018, Luderer_2021, Deangelo_2021}. The term IAM covers a wide variety of models including a global representation of energy, economy, climate and land. Generally, IAMs suffer from a low spatial and temporal resolution that prevents the representation of energy networks and the variability of wind and solar, although some efforts to improve these limitations are ongoing \cite{Luderer_2017, Pietzcker_2017, Brinkerink_2021}. On the other hand, Energy System Models (ESMs) with higher spatial and temporal resolution, as well as detailed representations of power networks \cite{Gils_2017, Plesmann_2017, Schlachtberger_2017, Horsch_2018b, Child_2019, Zappa_2019, Trondle_2020, Martinez-Gordon_2021} and storage \cite{Budischak_2013, Victoria_2019_storage, Cebulla_2017, Sepulveda_2021} are used to investigate power system transformations, but by focusing on regions and particular years, they miss long-term global interactions with climate and technological learning. Recently, some ESMs have been extended to include other sectors \cite{Brown_2018, Bogdanov_2019, Victoria_2020, Gea-bermudez_2021, Chang_2021} resulting in alternative transition paths that are characterized by a higher contribution from wind and solar photovoltaics (PV) than IAMs scenarios, together with high electrification of other sectors \cite{Victoria_2021, Jaxa-Rozen_2021}.

\

\begin{figure}[!h]
\centering
\includegraphics[width=\columnwidth]{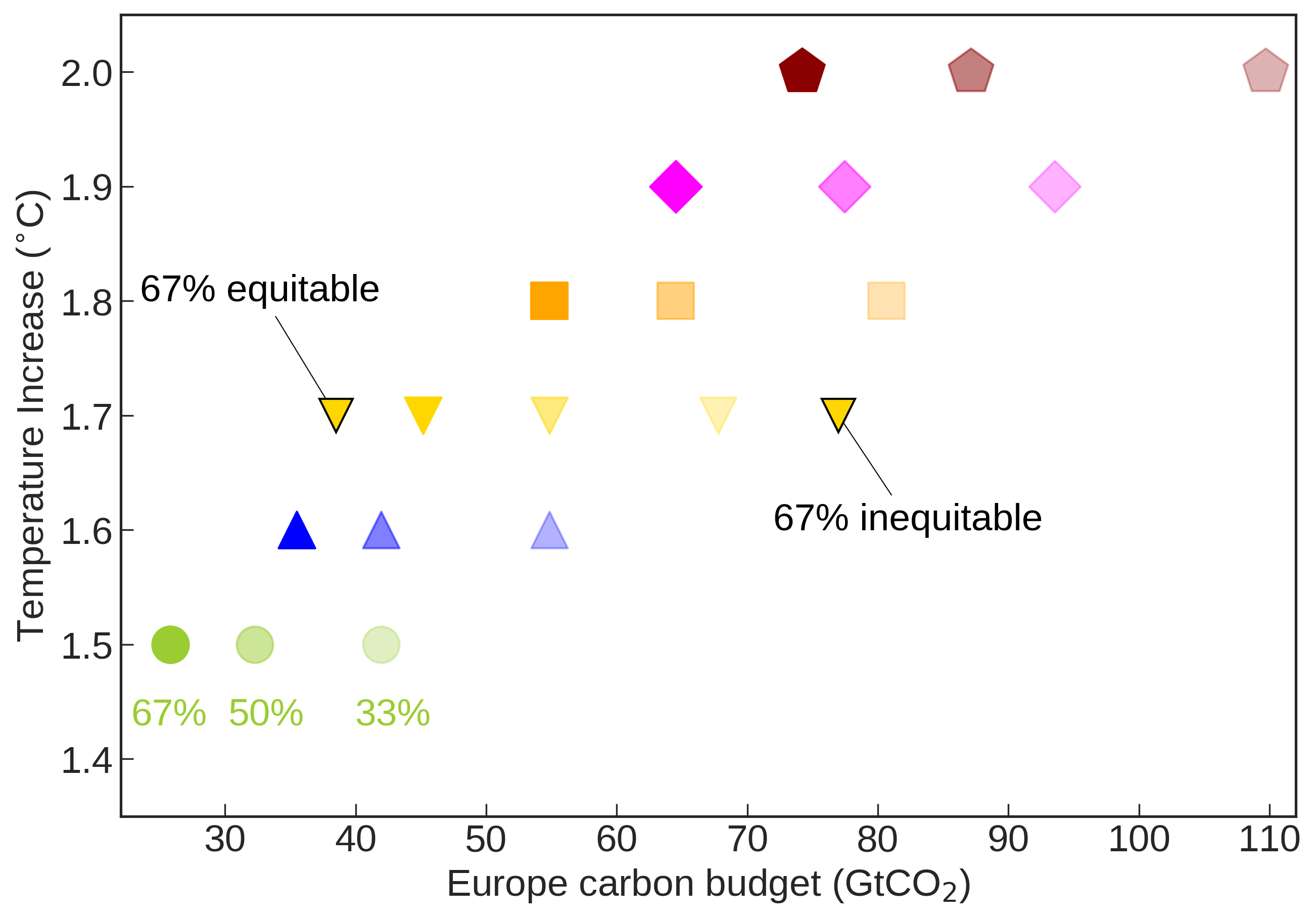}
\caption{Temperature increase for distinct carbon dioxide budgets in Europe from 2020 (6.4\% of global emissions allocated to Europe based on an equal per-capita distribution). Confidence intervals are indicated by different shadings. For 1.7$^{\circ}$C and 67\% confidence interval, the figure also shows the European budgets assuming a equitable distribution that compensates for historical emissions, and an inequitable distribution that assumes the prevalence of historical splits among regions (5.5\% \cite{Alcaraz_2018} and 11\% \cite{Raupach_2014} of global emissions allocated to Europe, respectively). } \label{fig_carbon_budget_vs_temperature} 
\end{figure}

Some of the authors have recently shown that, for a carbon budget corresponding to 1.75$^{\circ}$C temperature increase, it is less expensive to follow an early and steady decarbonisation path in Europe, in which emissions are strongly reduced before 2030, compared to delayed action that requires more abrupt and expensive transformations mid-century \cite{Victoria_2020}. Here, we go one step further by investigating the consequences of transforming the full sector-coupled European energy system with different cumulative emissions corresponding to various temperature increases. There are two main consequences of Europe achieving net-zero emissions while using a lower carbon budget. First, ceteris paribus, the probability to remain below a certain temperature increase is higher, or the associated temperature increase is lower, see Fig. \ref{fig_carbon_budget_vs_temperature}. Second, for the same global budget, reducing cumulative emissions in Europe enables higher emissions in other regions compensating for Europe's higher historical emissions.

\

\begin{figure}[!h]
\centering
\includegraphics[width=\columnwidth]{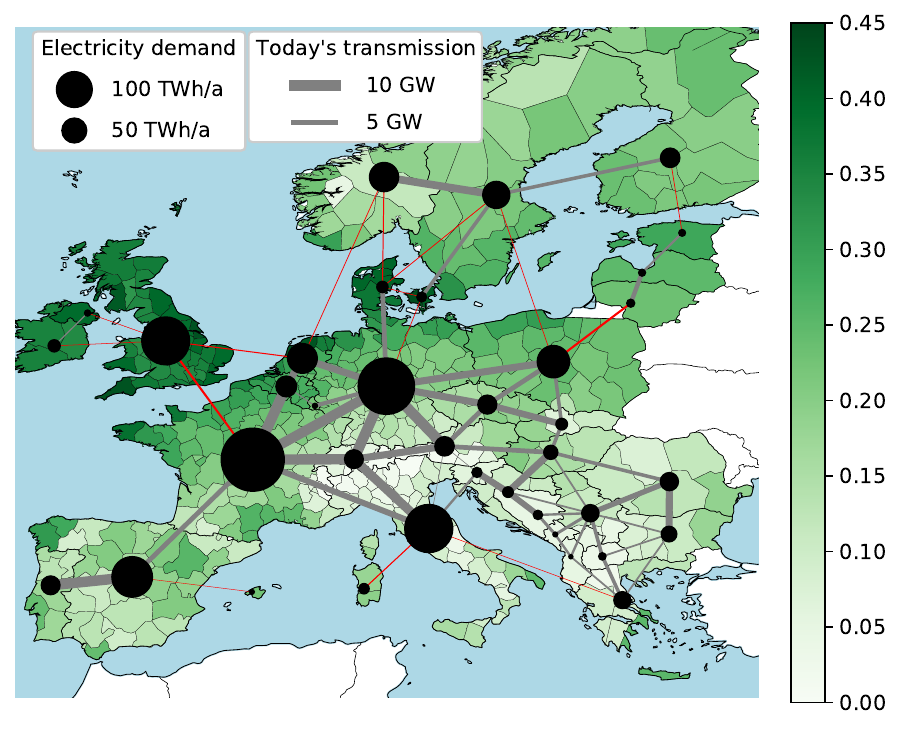}
\caption{The networked model comprises 37 nodes, one per region of countries belonging to separate synchronous zone. The size of the circles represents today's electricity demand in the residential and services sector. HVAC/HVDC transmission capacities among countries are shown in grey/red. Renewable resources are aggregated to the smaller regions shown on the map. The green shades represent the annual capacity factor for onshore wind in the different regions. Equivalent information for solar PV and offshore wind is depicted in Fig. S2-3. } \label{fig_map} 
\end{figure}

In this work, we use PyPSA-Eur-Sec\cite{pypsa-eur-sec}, an open model of the sector-coupled European energy system with uninterrupted 3-hourly resolution for a full year and a 37-nodes network, Fig. \ref{fig_map}. The model comprises the electricity, heating and land transport sectors used in \cite{Brown_2018, Victoria_2020}. Moreover, it is extended to include the transformation of industry, industrial feedstocks, shipping and aviation, the use of biomass and a detailed accounting of carbon capture, use, and storage (CCUS), as well as demand-side efficiency improvements in buildings. Our model includes higher time and spatial resolution than most IAMs. This captures the variability of wind and solar, the presence of heating demand peaks and dark doldrums (\textit{i.e.}, periods with low wind and solar generation), and the role of storage at different time scales. This, together with detailed modelling of electricity and hydrogen grids, is crucial to estimate flexibility needs to balance variable renewable generation. Moreover, we include a more detailed breakdown of industry by sector, including for example the option of direct reduced iron (DRI) in steel manufacturing. 

\

We model the transformation of the European energy system using a myopic approach in 5-years steps from 2020 to 2050 assuming different carbon budgets. This paper focuses on two main research questions which can be stated as: What are the economic consequences of different climate ambitions for Europe? When do key new technologies emerge under distinct budgets with a common set of cost and performance assumptions? 
We show that, regardless of the budget, similar technological transformations take place, but for the 1.5 and 1.6$^{\circ}$C budgets it is cost-optimal for most of them to take place already by 2035. We extend the existing literature by providing three main novelties. First, the use of a highly resolved model. Second, the comparative analysis of transition paths for Europe under carbon budgets corresponding to temperature increases discretised by a tenth of a degree. Third, the inclusion of a sensitivity analysis to the cost and potential of CO$_2$ sequestration.

\section{Methods}

\subsection{Baseline model setup}

We model the transformation of the European energy system under six different carbon budgets corresponding to a temperature increase between 1.5 and 2$^{\circ}$C, with 67\% confidence, see \cite{IPCC_1.5} and Supplemental Materials. The share of the global carbon budget allocated to Europe is estimated assuming an equal-per capita distribution \cite{Raupach_2014, Alcaraz_2018}. Fig.  \ref{fig_carbon_budget_vs_temperature} shows alternative relations between Europe carbon budget and temperature increase when (i) the historical responsibilities of every region are taken into account \cite{Alcaraz_2018}, (ii) splitting is proportional to historical emissions in every region \cite{Raupach_2014}, and (iii) different confidence intervals for temperature increase are considered \cite{IPCC_1.5}. 

\begin{figure}[!h]
	\centering
	\includegraphics[width=\columnwidth]{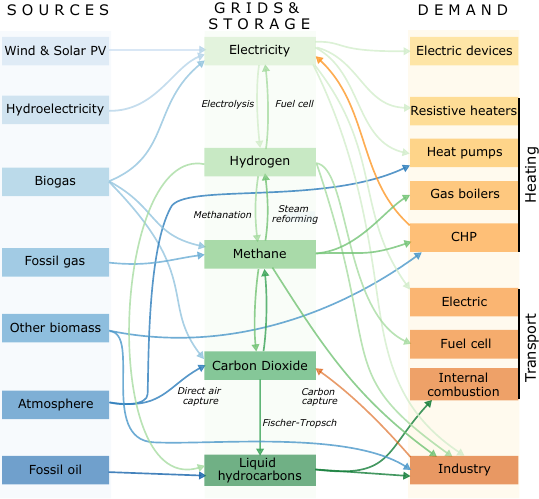}
	\caption{Possible energy flow paths from supply, via storage and transmission networks, to demand in the sector-coupled model PyPSA-Eur-Sec.} \label{fig_model_scheme} 
\end{figure}

The available carbon budget is distributed throughout the path assuming an exponential decay and forcing net-zero CO$_2$ emissions in 2050, Fig. S1. We impose net-zero CO$_2$ emissions in 2050 for all the paths to ensure that all of them are compatible with the committed EU target. Nevertheless, it is worth mentioning that the EU commitment of net zero-GHG in 2050 could require net negative CO$_2$ emissions by mid-century, depending on steps taken in the agriculture, forestry and land use sectors. A myopic approach is used in 5-year time steps from 2020 to 2050. In essence, for every time step, the total system cost is minimised subject to a CO$_2$ constraint determined by the exponential decay, but without any information regarding the future. Contrary to assuming perfect foresight, the myopic approach provides a more expensive solution. However, it captures better shortsighted decisions by policymakers and investors, and it is less impacted by the assumed social discount rate. Technologies installed in previous time steps remain in the system and contribute to the total system cost until they reach the end of their lifetime. For every time step, the open energy model PyPSA-Eur-Sec \cite{pypsa-eur-sec} including 37 nodes, network modelling and uninterrupted 3-hour resolution for a full year is used. Generation, storage, energy conversion, and transmission capacities are optimised assuming a long-term market equilibrium, as well perfect competition and foresight. PyPSA-Eur-Sec includes the electricity, heating, land transport, aviation, shipping, industry sectors and a detailed representation of the carbon cycle, Fig. \ref{fig_model_scheme}. 

\

While the use of uninterrupted hourly resolution has proven to be key to estimate the flexibility needs for highly renewable energy systems \cite{Victoria_2021, Brown_2018}, several analyses have proven that minor differences are found when using 3-hour instead of 1-hour time steps \cite{Pfenninger_2017, Schlachtberger_2018, Schyska_2021}. This is due to the fact that 3-hourly resolution is enough to capture the solar daily fluctuations and estimate the required short-term storage. Hence, we select 3-hour time steps to enable a high spatial resolution, network modelling and detailed technology description while reducing computational requirements. For ever year in one transition path, the model including 37-nodes network, 370 regions representing renewable resources and full sector coupled is solved in approximately 4 hours using 30GB RAM.

\

Exogenous assumptions in the model include the path of electrification of land transport, conversion of shipping to H$_2$ and some transformations in the industry. Each of these exogenous assumptions is discussed in more detail below. The model determines endogenously the technologies to produce electricity and heating, the origin of H$_2$ (electrolytic or by steam methane reforming with our without carbon capture), and the origin of methane and naphtha (fossil or synthetically produced) that are used in land transport and shipping, in aviation and for energy consumption and feedstock in the industry. The sectors are described in detail in the Supplemental Materials and a summary is provided below. 

\

Electricity can be produced by solar PV, onshore and offshore wind, open (OCGT) and combined-cycle gas turbines (CCGT), nuclear, coal, lignite power plants, and combined heat and power (CHP) units using biomass or gas. The solar and wind resource are represented by 370 regions, each of which is connected to one of the 37 nodes in the network. Electricity can be stored in batteries or H$_2$ storage (underground in salt caverns or overground in steel tanks). Reservoir, run-of-river hydro and pumped hydro storage (PHS) capacities are fixed exogenously based on existing facilities. The existing and planned transmission capacities are modelled using linear power flow. H$_2$ can be produced using electrolysers and by steam methane reforming (SMR) with or without carbon capture. A H$_2$ network can be built to connect countries if it is cost-effective. The impact of disallowing the H$_2$ network is evaluated in a sensitivity analysis. 

\

Heating demand can be supplied by heat pumps, heat resistors and gas boilers and stored in thermal energy storage. Costs and properties of these technologies vary depending on if they are installed in a high-density population area, where district heating systems are assumed, or in low-density population areas where only individual solutions are considered. In the former, heat can also be provided by CHP plants. Efficiency gains due to building retrofitting are exogenous, see Fig. S3 and \cite{Zeyen_2021}. The impact of assuming endogenous investment in building retrofitting is evaluated in a sensitivity analysis.

\

CO$_2$ can be captured from exhaust gases (CHP plants, SMR or process emissions in the industry) or by direct air capture (DAC). Captured CO$_2$ can be used to produced synthetic methane via the Sabatier reaction or synthetic hydrocarbons via the Fischer-Tropsch process. It can also be sequestered underground with a maximum potential of 200 MtCO$_2$/a, which is conservative but enough for capturing process emissions. Cost assumptions for different technologies are taken from DEA \cite{DEA_2019}. Future cost evolution of different technologies is exogenous to the model \cite{DEA_2019}, see Note S12. Efficiencies, lifetimes, and maximum potential for renewable technologies are described in the Supplemental Materials. Based on the JRC database \cite{JRC_biomass}, we follow a conservative approach in which only biomass that is not competing with crops is accounted as solid-biomass potential and can be used in the industry or burnt in CHP plants with our without carbon capture. A sensitivity analysis disallowing biomass with carbon capture is performed. Biogas is upgraded into biomethane. 

\

In the industry sector, the production of materials (such as steel, cement, chemicals) in every node is assumed to remain constant. A detailed analysis is carried out in every industrial subsector to model the most probable transformations. The general approach includes the electrification of some industrial process, and the use of methane and biomass for high and mid-temperature process heat, respectively. A comprehensive description of all the industrial transformations assumed is included in the Supplemental Materials. In every time step, the percentage of steel that is produced via direct reduced iron (DRI) is fixed exogenously and so is the supply of aluminium from scrap metal. The model determines endogenously how the hydrogen demand is supplied (either using electrolysers, SMR or SMR with CC) and whether the methane and hydrocarbons have fossil, synthetic or biogas origin. All hydrocarbon feedstocks are also accounted for in the model.

\

Road and rail transport transformation is exogenously fixed using a path that ends with 85\% of land transport electrified and 15\% using fuel cells in 2050. These assumptions are fixed exogenously because we expect consumer choice, government support and stock turnover inertia to be more decisive than pure cost optimisation. Half of the existing EVs in every time step are assumed to do smart charging and enable vehicle-to-grid operation. Shipping transformation is also exogenous and follows a path that entails full conversion to hydrogen in 2050. Aviation consumes kerosene whose origin (fossil vs. synthetic) is endogenously determined. The possibility of importing synthetic fuels to Europe is not modelled.
The agriculture sector is not included in the model and it is assumed that emissions from this sector are offset by the LULUCF sector. 

\subsection{Sensitivity analysis}
There are many uncertain input assumptions that flow into the optimisation model. Uncertain inputs could overwhelm any signal in the results. To explore how the results depend on the inputs, we perform several sensitivity analyses: varying the most important cost assumptions, leaving cost and technology assumptions fixed at 2020 values throughout the pathway, varying the carbon sequestration costs and volumes, endogenising building efficiency measures, removing the hydrogen network and excluding biomass with CCS.

\FloatBarrier

\section{Results}

\subsection{The timing of key technology transformations}
\begin{figure*}[!h]
\centering
\includegraphics[width=2\columnwidth]{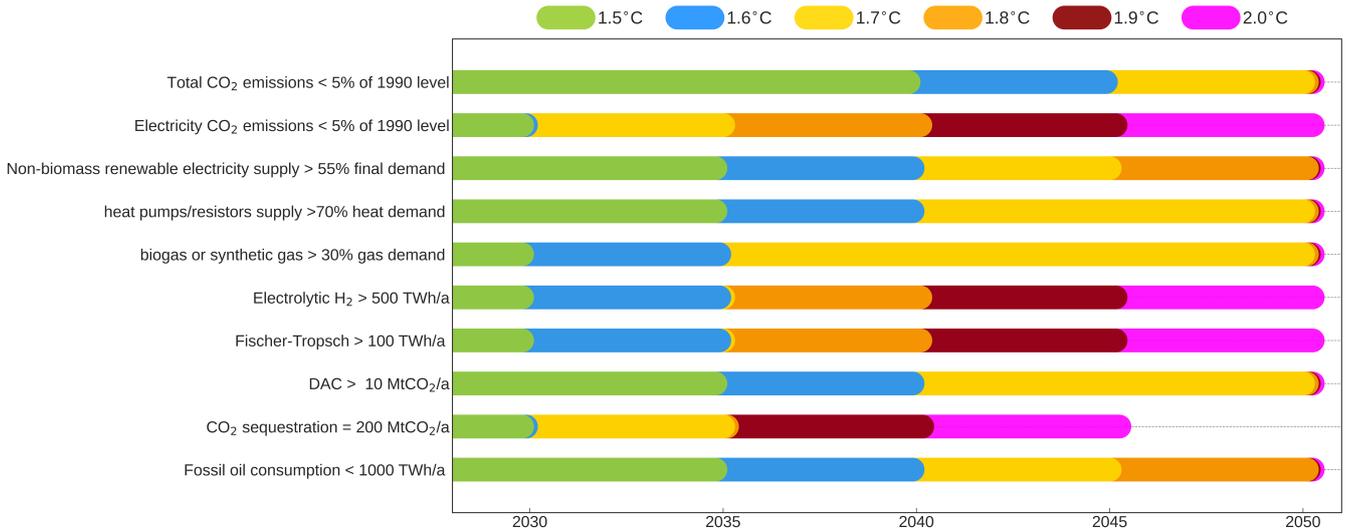}
\caption{\textbf{Occurrence of key transformations for distinct carbon budgets.} The bars indicate the period when key technological transformations (described by labels on the left) occur for modelled transition paths under carbon budgets corresponding to temperature increase between 1.5$^{\circ}$ and 2$^{\circ}$C. When the bar corresponding to a carbon budget is hidden by another color, this indicates that the timeline for the technological transformation is the same for both budgets. See Fig. S30, S34-44 for sensitivity analysis of these results.} \label{fig_KPI} 
\end{figure*}


\begin{figure}[!h]
\centering
\includegraphics[width=\columnwidth]{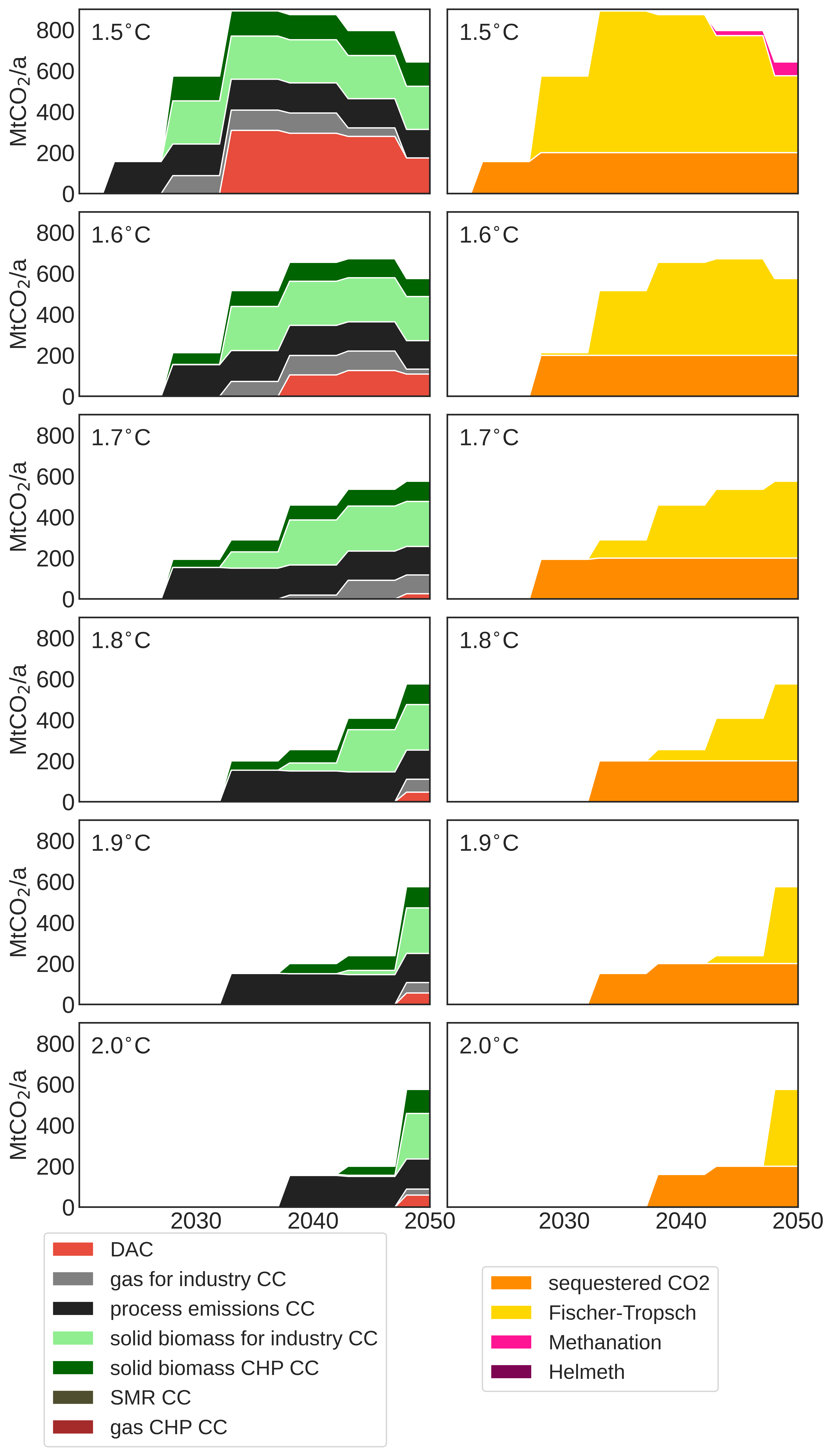}
\caption{(left) Technologies capturing CO$_2$ and (right) sequestration and use of CO$_2$ in the system for the distinct carbon budgets. CC stands for carbon capture. See Fig. S22, S25 for sensitivity analysis of these results.} \label{fig_NET} 
\end{figure}

Fig. \ref{fig_KPI} gathers the occurrence of some key transformations under distinct carbon budgets. Our scenario assumptions, which include exogenously determined CO$_2$ emissions paths, imply that lower carbon budgets reach net-zero emissions earlier, see also Fig. S1. The CO$_2$ emissions path corresponding to 1.5$^{\circ}$C carbon budget reduces emission in year 2040 to less than 5\% of 1990 level. The time step at which electricity generation is almost fully renewable precedes the system-wide decarbonisation by 5 to 15 years. For 1.5 and 1.6$^{\circ}$C-budgets, it would be optimal to fully decarbonise the electricity generation already in 2030, which highlights how challenging the low-emissions budgets are. Electricity generation is the first sector to undergo a deep transformation, see also Fig. S7. This was expected since renewable technologies, mainly solar PV, onshore and offshore wind are already competitive on a levelised basis with fossil-fueled generators \cite{WEO_2020, BNEF_2020}. Regardless of the budget, all the paths entail strong electrification of other sectors with non-biomass renewable electricity supplying more than 55\% of final energy demand in 2050. The pathways for 1.5 and 1.6$^{\circ}$C-budgets achieve more than 40\% of renewable primary energy in 2030, a target that has been recently proposed by the European Commission \cite{Fit-for-55}. However, those two budgets see a dramatic ramp-up of technologies, particularly solar PV, wind and electrolysis, between 2025 and 2035, see Fig. S5 and S6. In particular, for the 1.5$^{\circ}$C budget, the model install around 500 GW/a of new wind and solar in that period. The model finds it cost-effective to build a hydrogen network that enables exchanges among countries, Fig. S30. 

\

Our results show three strategies to decarbonise the heating sector. First, heat pumps and electric resistors are used to supply heating demand. Second, in urban areas, district heating systems enable the use of large heat pumps together with biomass CHP units and gas boilers to ensure heating supply in the winter. When the system approaches net-zero emissions, waste heat output from the Fischer-Tropsch process dumped into district heating systems can cover up to 20\% of the demand in those areas. Third, in regions without district heating systems, gas boilers are used at peak demand times to backup the heat pumps that cover the main supply. The share of consumed gas that has fossil, biogas or synthetic origin evolves throughout the transition paths, Fig. S17. The share of technologies providing heat in every time step are shown in Fig. S11-15.

\

Between 2030 and 2040, substantial production of electrolytic H$_2$ is cost-effective for scenarios below 1.9$^{\circ}$C. Initially, the production H$_2$ of is used for seasonal balancing of power generation, see Fig S32. As the Fischer-Tropsch technology is installed, H$_2$ is consumed to produce synthetic hydrocarbons reaching a demand of 1700 TWh$_{H2}$/a when all the hydrocarbons consumed in the model are synthetically produced. None of the transition paths installs capacity to produce blue H$_2$ via steam methane reforming with carbon capture (SMR-CC), see Fig. S25. The production of H$_2$ switches straight from SMR to electrolysis. The production of synthetic hydrocarbons starts as soon as the electricity generation is fully decarbonised in every budget. 

\

Direct air capture (DAC) capacity is only installed for carbon budgets corresponding to temperature increase below 1.7$^{\circ}$C. The model finds it more cost-effective to capture CO$_2$ from (i) process emissions (which in 2050 represents 155MtCO$_2$/a, Note S10), (ii) biomass and methane used in the industry, and (iii) biomass combusted in CHP units, see also Fig. \ref{fig_NET}. With the reference cost assumptions, sequestering CO$_2$ underground is economically preferable and it occurs earlier than building Fischer-Tropsch capacities that transform the captured CO$_2$ into synthetic hydrocarbons. The 200MtCO$_2$/a potential for CO$_2$ sequestration assumed for Europe is fully utilised as soon as it becomes cost-effective. A sensitivity analysis for CO$_2$ sequestration is conducted in Section \ref{sec_CO2_storage_sensitivity}. 

\

The CO$_2$ emissions paths for the 1.5 and 1.6 $^{\circ}$C budgets are assumed to reduce emissions below 5\% of 1990 by 2040 and 2045 respectively. However, the exogenously defined transformation of land transport and shipping still include substantial emissions at that time. This result in the model including a large deployment of Negative Emissions Technologies (NETs) between 2030 and 2045 for those scenarios. This example illustrates the consequences for the system if the decarbonisation of some sectors lags behind the global CO$_2$ reduction targets. 

\

In 2015, Europe imported 6000 TWh/a of oil \cite{TYNDP_scenarios}. More stringent carbon budgets reduce Europe's external dependency earlier. The reasons are twofold: first, efficiency measurements and direct-electrification reduces the demand for oil and methane; second, as CO$_2$ emissions are constrained, the upgrade of biogas into methane and the production of synthetic oil become cost-effective, see Fig. S17-S18.

\subsection{Sectoral emissions and CO$_2$ price} \label{sec_co2_price}
In all the alternative transition paths, the order of sectoral emissions reductions is maintained, see Fig. S8. Electricity generation is decarbonised first, followed by the heating and industry sectors, and finally, aviation. In our analyses, the shares of road transport and shipping that gets electrified or transformed into using hydrogen are exogenously fixed, Note S11.

\begin{figure}[!h]
\centering
\includegraphics[width=\columnwidth]{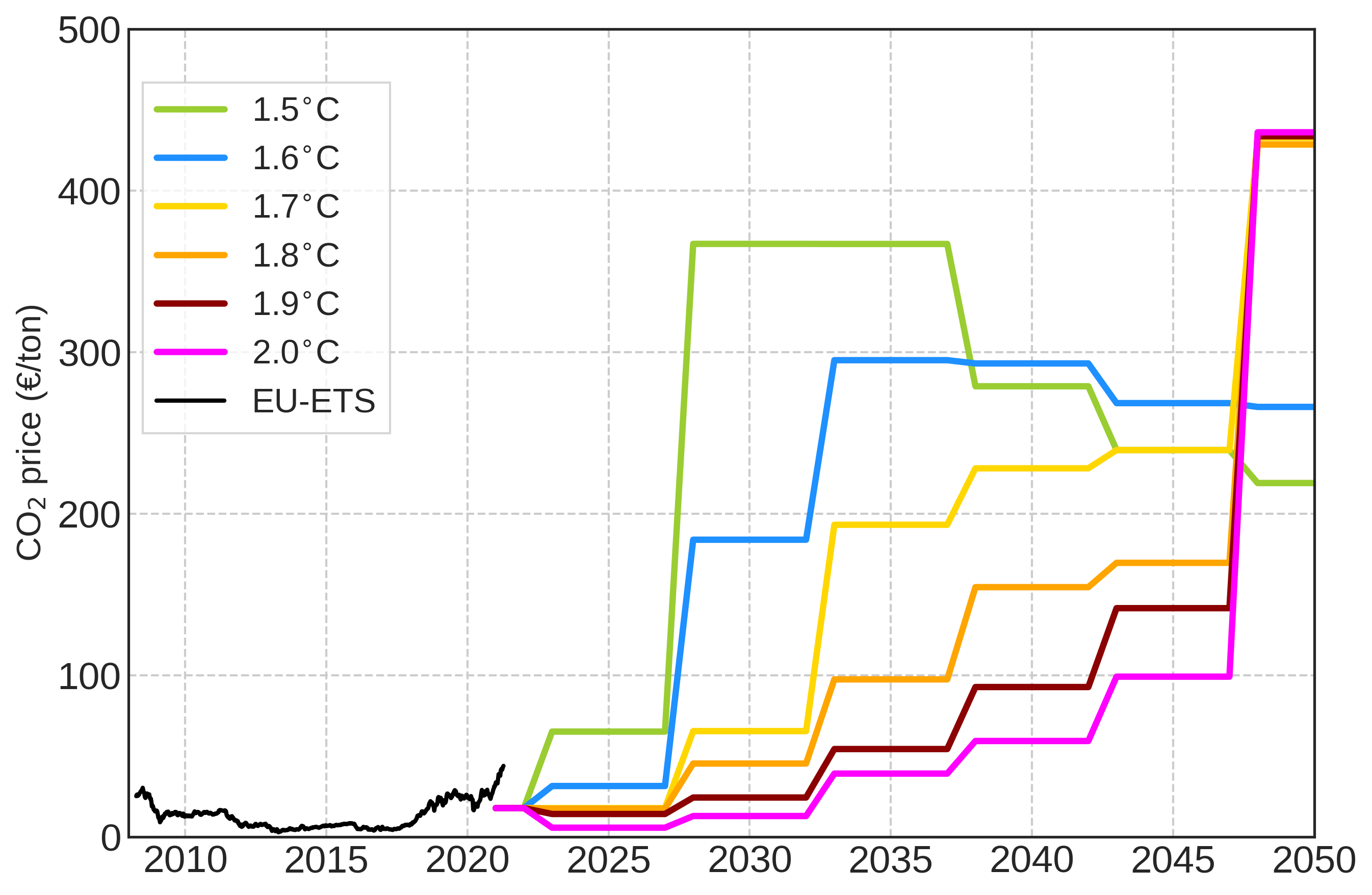}
\caption{Required CO$_2$ price for distinct carbon budgets. The CO$_2$ price is not an input to the model, but an output calculated as the Lagrange/KKT multiplier associated with the CO$_2$ cap constraint (Eq. 13 in Supplemental Materials). See Fig. S32 for sensitivity analysis of these results.} \label{fig_CO2_price} 
\end{figure}

\

The required CO$_2$ price, also known as the marginal abatement cost, is an output of the model. It increases as CO$_2$ emissions allowance are reduced, see Fig. \ref{fig_CO2_price}. For the 1.5$^{\circ}$C-budget, a sharp increase of the CO$_2$ price, reaching 370 \EUR/tCO$_2$, is required to incentivise the extremely fast build-up of a carbon-neutral system by 2035.  The 1.6$^{\circ}$C carbon budget requires a smoother ramp-up in CO$_2$ price that stabilizes towards the end of the transition at around 270 \EUR/tCO$_2$. Higher carbon budgets require an increase of CO$_2$ price in 2050 to force carbon neutrality. Compared to our previous analysis \cite{Victoria_2020}, we found that similar CO$_2$ prices are required by mid-century, even though in this work we included a broader representation in the model of NETs such as carbon capture in the industry and CHP units, carbon sequestration, and carbon use in the Fischer-Tropsch process. The assumed cost and potential of CO$_2$ sequestration also affect the required CO$_2$ price, as discussed in Section \ref{sec_CO2_storage_sensitivity}. It is important to realize that, by setting up a CO$_2$ cap in every time step, instead of assuming a CO$_2$ price that steadily increases throughout the transition, the model avoids the large use of carbon dioxide removal as discussed by Strefler \textit{et al. }\cite{Strefler_2021}.

\subsection{Net-present-value of system costs}

\begin{figure*}[!h]
\centering
\includegraphics[width=\textwidth]{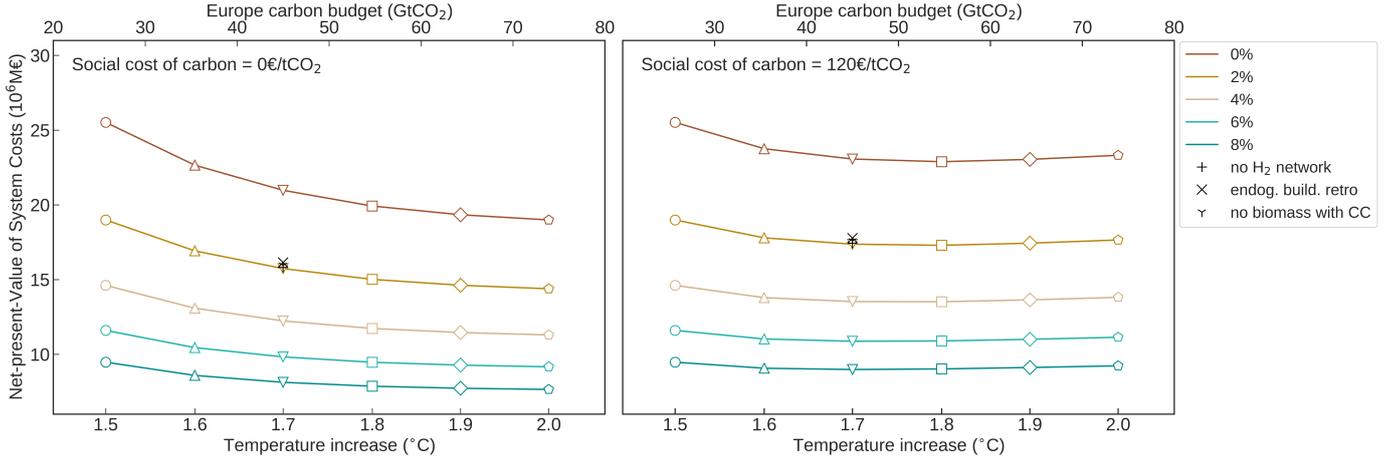}
\caption{Net-present-value of system costs for different carbon budgets and discount rates (indicated in the legend). (left) No social cost of carbon included. (right) A social cost of carbon of 120\EUR/tCO$_2$ is assumed. Equivalent results for other social costs of carbon are shown in Fig. S9 and S10.} \label{fig_cumulative_system_cost} 
\end{figure*}

Figure \ref{fig_cumulative_system_cost} depicts the net-present-value (NPV) of system costs for transition paths using distinct carbon budgets and social discount rates. The assumed social discount rate has a higher impact on the calculation than the carbon budget in every path. As expected, the net-present-value of future costs is lower with higher discount rates. Fig. \ref{fig_cumulative_system_cost} also shows that the required investments for distinct climate ambitions are not that different. Lower budgets require an earlier build-up of new assets, which due to the exogenous evolution of costs assumed, results in a more expensive transition. For a detailed description of cost components throughout the system transformation see Fig. S4. 

\

For a 2\% discount rate, the 1.5 and 1.6$^{\circ}$C budgets are, respectively, 8\% and 1\% more expensive than the 2$^{\circ}$C budget. These percentages are 5\% and -1\% when a 6\% discount rate is used. The implications of the assumed discount rate have been extensively discussed by other authors, see \cite{Garcia-Gusano_2016, Emmerling_2019, Hermelink_2015} and Supplemental Materials. For transition paths with perfect foresight and negative emissions, Emmerling \textit{et al.} \cite{Emmerling_2019} found that reducing the discount rate from 5\% to 2\% more than doubles the required CO$_2$ price in 2020, more than halves the carbon budget overshoot and increases substantially the investments in renewable energy. In our analysis, the assumed discount rate does not change the timing of different technologies because we use a myopic approach in which the CO$_2$ emissions paths are set exogenously and not by imposing a single carbon budget constraint. Still,  this parameter has a large influence when computing the net-present-value of system costs. The selection of a discount rate higher than zero is based on the assumption that economic growth will continue. Based on this, historical growth rates are typically assumed. Temporal preferences are also claimed as an argument to support high discount rates. However, it is important to recognize the impacts in terms of inter-generational burden-sharing of this argument, as mitigation and adaptation costs in the short and long term will be paid by different groups of people. 

\

\subsection{Including climate damage costs}

So far we have considered only the cost of mitigation given defined CO$_2$ budgets. Including the costs saved by avoided climate damages could favour scenarios with tighter budgets, but there is a high uncertainty associated with the assessment of the economic impact of climate change \cite{planetary_economics, Tol_2013, Havranek_2015, Wang_2019, Pizer_2014, Guivarch_2016, Nordhaus_2017}. In this section we illustrate the potential impact on our results of including different assessments of climate damages quantified through the social cost of carbon (SCC). The costs of mitigation for tighter CO$_2$ budgets are compared with the reduced climate damage costs.

The social cost of carbon (SCC) represents the economic cost caused by an additional tonne of carbon dioxide emissions or its equivalent. SCC is different from the marginal abatement cost discussed in Section \ref{sec_co2_price}. On the one hand, SCC is typically employed in cost-benefit analysis, aiming at evaluating to what extent the cost of climate mitigation compensates for the avoided climate change impacts. On the other hand, the CO$_2$ marginal abatement cost is employed in cost-effectiveness analyses, \textit{i.e.}, when estimating the most cost-effective strategy to attain a climate target. Arguments in favor of using marginal abatement cost \cite{Stern_2021} and SCC \cite{Aldy_2021} have been put forward, and some authors argue for using a welfare-optimal carbon price calculated as the sum of both \cite{Schultes_2021}.

\

In our analysis, the marginal abatement cost is an output of the model, determined as the Lagrange/KKT multiplier of the emissions cap constraint imposed every year, which in turn has been exogenously determined to make sure that the carbon budget is not exceeded. For instance, for the 2.0$^{\circ}$C transition path, the marginal abatement cost takes care of ensuring that cumulative emissions correspond to a temperature increase of 2.0$^{\circ}$C. The marginal abatement cost depends on the transition path and the year, Fig. \ref{fig_CO2_price}.  
 
\

On top of that, when we compare the NPV of system costs for different carbon budgets, we want to include the fact that the 2$^{\circ}$C budget is expected to have stronger economic impacts caused by climate change than the 1.5$^{\circ}$C budget. To take that into account, when estimating the NPV, we add a term that is calculated as the SCC multiplied by the 
additional emissions of every transition relative to the 1.5$^{\circ}$C-budget. In this way, we limit warming to a temperature threshold (different for every carbon budget) using the marginal abatement costs, and also account for the damages occurring below that threshold via the SCC.

\
 
Widespread SCC estimations can be found in the literature caused by the uncertainties associated with climate change and duration of impacts, together with the high sensitivity to some modelling assumptions \cite{planetary_economics, Tol_2013, Havranek_2015, Wang_2019, Pizer_2014, Guivarch_2016, Nordhaus_2017}.  We have considered here a SCC of 120 \EUR/tCO$_2$ \cite{SCC_OECD}. For simplicity and easy comparison of the scenarios, we assume that the SCC is constant in time. Fig. S9 and S10 show, respectively, the cumulative system cost for a SCC of 75 and 300 \EUR/tCO$_2$ and we use the same SCC regardless of the discount rate used in our calculations, despite the fact that the SCC itself depends on the discount rate. The former represents the recommendations by the Danish Economic Council of Environmental Economics \cite{DECEE2019}. For the latter, the transition under 1.5$^{\circ}$C already results less expensive than for the 2$^{\circ}$C budget. It must be noted that a SCC of 300 \EUR/tCO$_2$ is lower than the SCC estimated by the German Environmental Agency with 0\% time preference (680 \EUR/tCO$_2$) \cite{SCC_BAU} and similar to recent recommendations by the European Commission (250 and 800 \EUR/tCO$_2$ in 2030 and 2050 respectively) \cite{SCC_EU}. Moreover, considering the risk associated with climate tipping points could increase the SCC by up to a factor of two \cite{Dietz_2021}. On top of that, the co-benefits of reducing CO$_2$ emissions in Europe due to avoided premature mortality and morbidity caused by air pollution, reduced lost workdays and increased crop yields are estimated in the range of 125–425 \EUR/tCO$_2$ \cite{Vandyck_2018}.

\

We believe that Fig. \ref{fig_cumulative_system_cost} provides a useful decision map where the costs of distinct climate ambitions in Europe are compared including not only those cost components associated with a profound transformation of our energy system but also those related to the avoided CO$_2$ emissions. Together with Fig. \ref{fig_carbon_budget_vs_temperature}, it enables comparing the alternative budgets related to different temperature increases, confidence intervals and Europe's share of global emissions, while not losing sight of the strong impacts of exogenous assumptions such as the discount rate or the SCC.

\subsection{Sensitivity 1: Costs and technological performance} \label{sec_costs_sensitivity} 

The sensitivity to some of the main assumptions is evaluated through three sets of analyses. First, the sensitivity to cost assumptions for selected technologies is described in the Supplemental Materials. Assuming variations of $\pm$20\% of the costs of the main technologies do not have any substantial impact on the date for key transformations, Fig. S34-S41. We found the difference in NPV of system cost among different budgets, the use of solar and wind as main energy generators, and the selection of electrolytic-H$_2$ over the production via SMR-CC to be robust for all the cost variations, Fig. S29.

\ 

Moreover, to evaluate the impact of the assumed exogenous cost evolution, a sensitivity run is implemented in which costs and technologies performance are kept fixed at values corresponding to 2020. This produces no major impact on the timeline for key transformations, Fig. S30, the use of NETs, Fig. S31 and the production of H$_2$ and synthetic fuels, Fig. S29. However, neglecting the potential costs evolution translates into a higher contribution from wind and nuclear compensating lower solar generation, Fig. S29, higher CO$_2$ prices, Fig. S32, higher NPV of system costs, Fig. S33, and slightly higher cost differences between the transition paths with lower carbon budgets and 2.0$^{\circ}$C, Fig. S29.

\subsection{Sensitivity 2: CO$_2$ sequestration cost and potential} \label{sec_CO2_storage_sensitivity} 

\begin{figure}[!h]
\centering
\includegraphics[width=\columnwidth]{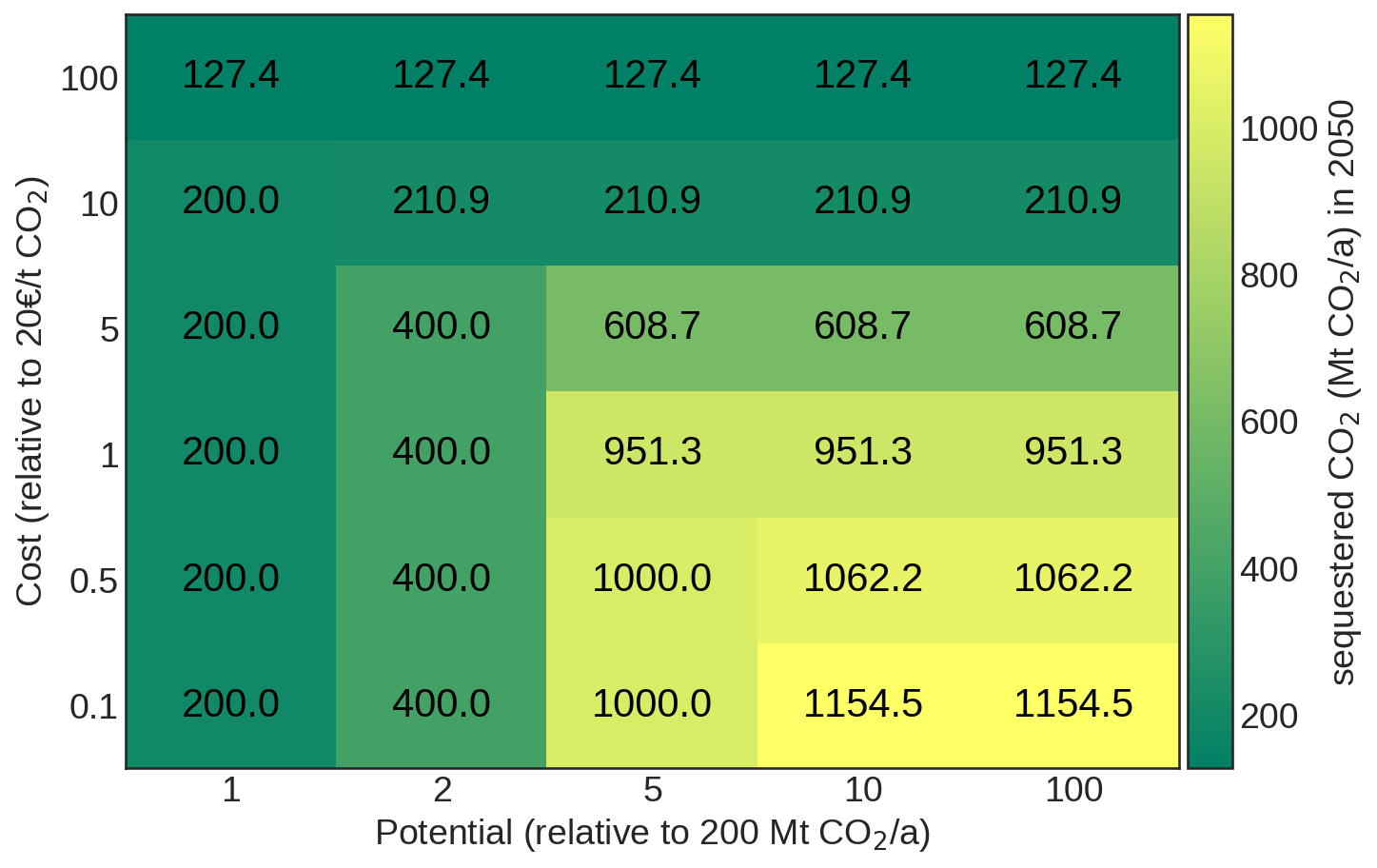}
\caption{Sensitivity Analysis: CO$_2$ sequestered underground per year at the end of the transition for the 1.7$^{\circ}$C-budget. Results are shown as a function of the assumed CO$_2$ sequestration cost and potential.} \label{fig_sensitivity_cc_co2_stored} 
\end{figure}

So far, we have assumed a CO$_2$ sequestration potential for Europe of 200 MtCO$_2$/a, and a cost for sequestration and transport of 20\EUR/tCO$_2$. The high uncertainties associated with those assumptions and the full deployment of the potential shown in Fig. \ref{fig_NET} motivates a sensitivity analysis specifically focused on CO$_2$ sequestration. We conduct it here for the 1.7$^{\circ}$C-budget. Fig. \ref{fig_sensitivity_cc_co2_stored} shows the amount of CO$_2$ sequestered in 2050 under different assumptions. As the potential is increased, the system finds it optimal to sequester up to around 950MtCO$_2$/a. It is important to realized that this value is far below the technical potential for CO$_2$ sequestration in Europe, estimated at 126 GtCO$_2$/a \cite{CO2_storage_potential}. This could reduce the annualised system cost by 10\%, and it would also impact the required CO$_2$ price, see Fig. S19 and S20. 

\

Besides underground sequestering, the alternative uses of captured CO$_2$ in the model are to produce synthetic methane or synthetic oil. The former is not part of the solution for most of the sensitivity cases, see Fig. S21, so we can consider the absence of methanation in the optimal system a robust result. Conversely, the production of synthetic liquid fuels strongly depends on the CO$_2$ underground sequestering potential, Fig. S21, S23. In short, when underground sequestering is hampered by high cost or limited potential, more synthetic fuel is produced via Fischer-Tropsch, as this is the second most cost-effective option. 

\

When lower costs are assumed for CO$_2$ sequestration, the rate increases slightly up to 1,150MtCO$_2$/a, still much below the estimated physical potential. It is also interesting to realize that, when the model selects a high CO$_2$ sequestration rate, additional CO$_2$ capture technologies like gas CHP with carbon capture are installed, although this technology was not part of the optimal solution under the reference potential of 200 MtCO$_2$/a, Fig. S23. Hydrogen production via steam methane reforming with carbon capture (SMR CC) is only part of the solution when the sequestration potential is extended to 1000 MtCO$_2$/a, Fig. S22-S23. Even with a extremely low cost assumption for CO$_2$ sequestration (2 \EUR/tCO$_2$) the use of SMR CC is limited to 220 TWh/a. 

\

If we look now at the opposite scenario, that is, one where the transport and sequestration of CO$_2$ ends up being more expensive than initially estimated, this results in a lower optimal sequestration rate, \textit{e.g.}, for a hundredfold increase in cost, the maximum CO$_2$ sequestration rate is approximately 127MtCO$_2$/a. As previously mentioned, the lower CO$_2$ sequestration rate is compensated by a higher production of synthetic oil, Fig S21.

\subsection{Sensitivity 3: H$_2$ network, building retrofitting, biomass with carbon capture} 

In this section, we briefly evaluate some additional assumptions. We start by discussing the constraints on the networks expansion. The previous analysis assumed no expansion of transmission links among countries, besides those already existing. The build-up of a greenfield hydrogen network among countries, which was cost-effective in all the transition paths, was also included, see Fig. S24. On the one side, for the 1.7$^{\circ}$C-budget, allowing the expansion of the power grid up to twice today's volume reduces the system cumulative cost by 2\%. As previously shown, the cost benefits of grid expansion \cite{Schlachtberger_2017} are reduced when the additional local flexibility provided by sector coupling is included in the model \cite{Brown_2018}. On the other side, disabling the build-up of the hydrogen network (\textit{i.e.} hydrogen is produced and consumed locally) only increases the cumulative system cost by 0.5\% and does not result in substantial changes on the key transformation indicators, Fig. S42. 

\

Previous analyses assumed exogenous reduction in space heating due to building retrofitting whose cost is not included. To evaluate this assumption, we implemented a sensitivity run where the extension of investment in building retrofitting is endogenously calculated as described in \cite{Zeyen_2021}. For the 1.7 $^{\circ}$C, this strategy is gradually implemented throughout the path as it becomes cost-effective, see Fig. S28.

\

CO$_2$ capture on biomass burnt in CHP plants or the industry sector shows substantial deployment when the system approaches net-zero emissions, Fig \ref{fig_NET}, but this technology is controversial due to the uncertain costs, land use and environmental impacts \cite{Gambhir_2019, Creutzig_2019, Vanvuuren_2018} . We run a sensitivity analysis disallowing CO$_2$ capture from biomass. The cumulative system cost is roughly the same, but the system deploys direct air capture to achieve the required negative emissions, Fig. S25 and S44.

\subsection{Limitations of this analysis}
Before concluding, we briefly mention the main limitations of our study in this section. First, we have assumed an exogenous transformation of land transport and shipping. Regarding land transport, we expect stock turnover, government support as well as public perception of new technology to be a stronger determinant of new vehicle shares that pure cost optimisation. Regarding shipping, we assumed a conservative transition since there is inertia in stock turnover that limit the transformation speed. This has important implications, particularly for the 1.5$^{\circ}$ and 1.6$^{\circ}$C budgets, which require offsetting the emissions from these sectors when net-zero emissions are imposed. 

\

Second, we have assumed that the cost evolution of different technologies is exogenous based on pathways forecast by the Danish Energy Agency, \textit{i.e.}, no endogenous learning effects are considered. By doing that, we are assuming that global learning, driven by globally installed capacities, will determine the future costs of technologies, but we do not represent possible local learning. We have evaluated the implications of the assumed costs evolution via sensitivity analyses. 

\

Detailed analyses of the near-optimum solution space for the European power system have been recently presented \cite{Neumann_2021, Pedersen_2021, Neumann_2021b, Sasse_2020}. In general terms, they found that the optimal solution space is quite ``flat'', meaning  that alternative solutions, in which the deployment of some technologies is different, can still achieve a cost close to the minimum. In this paper, we have not investigated the near-optimal solutions for the alternative transition paths, which could allow certain technology transitions to take place earlier or later. This remains a topic for further research.

\FloatBarrier

\subsection{Results Summary}

In this work, we have investigated the transformation of the European energy system between 2020 and 2050 under distinct carbon budgets corresponding to various temperature increases between 1.5$^{\circ}$ and 2$^{\circ}$C. We found that all the transition paths experience similar technological transformations, but they occur at different points in the future. 
The system begins decarbonising electricity generation by installing solar PV, onshore and offshore wind capacities.  This triggers strong electrification of other sectors such as heating, where large capacities of heat pumps are installed. Renewable electricity is also used to produce hydrogen via electrolysis, displacing the current production via steam methane reforming (SMR). 
A hydrogen network interconnecting the countries, co-optimised with the rest of the system, appear after 2035. Only for CO$_2$ sequestration potential higher than 1000 MtCO$_2$/a and low cost, the model installs limited capacities of SMR with carbon capture, casting doubts on the relevance, from a system perspective, on the blue hydrogen strategy. When the system approaches net-zero emissions,  electricity is also used to produced synthetic oil via the Fischer-Tropsch process, enabling the decarbonisation of the aviation and industry sectors. Carbon budgets corresponding to 1.5$^{\circ}$ and 1.6$^{\circ}$C result in cost-optimal solutions where the most substantial technological transformations are fully accomplished by 2030, which shows how challenging these budgets are.

\

The system installs negative emission technologies (NETs) to offset process emissions from the industry. NETs are also needed when the transformations of land and maritime transport lag behind the CO$_2$ reduction targets. First and foremost, CO$_2$ is captured from point-source emitters including process emissions and biomass burnt in the industry and CHP units. For the 1.5 and 1.6$^{\circ}$C-budget, direct air captured is also extensively used. The 200 MtCO$_2$/a assumed as CO$_2$ sequestration potential is completely deployed when the system approaches net-zero emissions. After that, the Fischer-Tropsch process is used to convert CO$_2$ into synthetic fuel.  The impacts of uncertainties in CO$_2$ cost and potential have been investigated via sensitivity analysis. For optimistic cost and potential assumptions, up to 1,150 Mt CO$_2$/a are used, which is far below the estimated physical potential of CO$_2$ sequestration in Europe (126 GtCO$_2$/a) \cite{CO2_storage_potential}. When the cost of CO$_2$ transport and sequestration is assumed to be higher, the use of CO$_2$ sequestration is reduced, and the full potential is not deployed. It is relevant to notice that under extremely high-cost assumptions (2,000 \EUR/tCO$_2$), it is still cost effective to sequester 127 MtCO$_2$/a. This partially compensates for the industry process emissions, which account for 155 MtCO$_2$/a in 2050, and results in a marginal abatement cost of approximately 2,100 \EUR/tCO$_2$. Hence, our results indicate that the lack of a small potential of CO$_2$ sequestration at a reasonable cost makes the net-zero CO$_2$ system very challenging.

\section{Discussion}

After presenting the main results of our analysis and its limitations, we discuss three main implications for the transformation of the European energy system.

\

\textbf{The need for high CO$_2$ prices.} Although only large emitters are currently included in the European Emissions Trading System (ETS), extending this mechanism to other sectors has also been proposed \cite{Edenhofer_2021, Fit-for-55}. Based on our results, for low-carbon budgets, ETS-sectors tend to reduce emissions earlier than non-ETS sectors, see Fig. S7. In particular, reducing emissions in the heating sector is expensive, mainly due to the strong seasonality in heating demand, see Fig. S27, and the fact that only in district heating systems (and not in individual systems) seasonal storage can be used for balancing. The need for stronger incentives to reduce emissions in the heating and industry sector is indicated in our results by the high CO$_2$ price required towards the end of the transition paths, Fig \ref{fig_CO2_price}.

\

Having a single CO$_2$ price in Europe for all the sectors and countries can create tensions. For instance, the ETS CO$_2$ price has reached values higher than 80\EUR/tCO$_2$ in 2021. For the electricity sector, the substitution of coal by gas power plants was already triggered at much lower CO$_2$ prices. The CO$_2$ price surge has increased the marginal cost of gas, which together with the rise in gas price, triggered an unprecedented period of high electricity prices in Europe. This has triggered some discussion on the negative effect of such high CO$_2$ prices for the electricity sector, but we can see that historical values are still much lower than those required in our model when approaching net-zero CO$_2$ emissions. This is one of the examples of the challenges for policy to reconcile the use of a single CO$_2$ price for sectors that required very different policy pressure. Similarly, a certain CO$_2$ price can represent a strong incentive for some European countries while not being strong enough to incentivize transformations in others \cite{Schwenk-Nebbe_2021, Pedersen_2022}.  Although the need for external action is indicated in our model by high CO$_2$ prices, in reality, high CO$_2$ prices may lead to negative distributional impacts and uncertainty if prices are volatile. Other policy implementations may be preferred, such as subsidies for emissions-free technologies and transformation deadlines including banning of emitting technologies.

\

\textbf{Large wind and solar PV deployment together with negative emissions technologies.} Our results show, consistently for all the carbon budgets and sensitivities, a strong deployment of wind and solar PV that become the cornerstone of the energy supply. This confirms previous works that emphasize the need for using up-to-date costs assumptions for wind and solar and proper modelling of balancing strategies to avoid downplaying this decarbonisation strategy \cite{Creutzig_2017, Victoria_2021, Luderer_2021}.

\

We found that at least a small amount of CO$_2$ sequestration, in the order of 200 MtCO$_2$/a, at a reasonable price, is needed to ease the operation of the fully decarbonized system. Nevertheless, we found that that it is possible to attain net-zero CO$_2$ emissions while using conservative assumptions for biomass potential and CO$_2$ sequestration. In essence, by avoiding common assumptions that are known to favour NETs we found cost-effective solutions that make limited use of them. The avoided pitfalls include (i) a poor representation of the balancing options and/or costs of renewable energy sources \cite{Victoria_2021}, (ii) perfect foresight optimisation with high discount rates which is known to result in large contribution from NETs \cite{Emmerling_2019, Mclaren_2020}, (iii) assumptions of an exogenous exponential increase of the CO$_2$ price \cite{Strefler_2021}. Conversely, in our model, we use a myopic approach and set up an emissions cap in every time step.

\

\textbf{Low carbon budgets are economically beneficial but require a dramatic increase in wind and solar capacities.}
When comparing alternative climate ambitions, the cumulative system cost is typically employed to assess the alternatives. However, the high impact of uncertain exogenous assumptions, such as the social cost of carbon (SCC) or the discount rate, on those estimations requires careful evaluation. Here, we found that for a 2\% discount rate and SCC of 120 \EUR/tCO$_2$, the 1.5$^{\circ}$ and 1.6$^{\circ}$C-budgets are 8\% and 1\% more expensive respectively, relative to the 2$^{\circ}$C budget. However, for a SCC of 300 \EUR/tCO$_2$, the 1.5$^{\circ}$ budget is cost-optimal.  In essence, our results show that the main challenge for low-carbon budgets is not the increase in the investments but the required dramatic ramp-ups of technologies, particularly solar PV, wind and electrolysis, between 2025 and 2035, see Fig. S5. In particular, for the 1.5$^{\circ}$C budget, the model install around 500 GW/a of new wind and solar in that period.

\

It is difficult to assess whether this dramatic development will be feasible in light of historical building rates. Technological transitions are known to be nonlinear and are better described by S-curves \cite{Marchetti_1979, Grubb_2020}. Cherp and co-authors \cite{Cherp_2021} evaluated historical data on wind and solar deployment in different countries and, by using an S-curve approximation, concluded that the maximum annual rate is expected to be 0.9\% and 1.1\% of the electricity supply for solar and onshore wind respectively. For the current EU annual electricity consumption of 2850 TWh/a, and assuming 1,200 and 2,000 full-load hours for wind and solar respectively, this translates into 35 GW/a, a much lower figure than the requirements in our model.  However, as argued by other authors \cite{Lowe_2022, Grubb_2020}, it is likely that the historical development rates, which were highly determined by policy support are not a good indicator of the feasible development speed for wind and solar once these technologies have become competitive and in many locations the lowest-cost option.

\

As a final remark, our simulations suggest there is only a small cost to ambitious emissions reduction. In this case, following the precautionary principle and pursuing an ambitious path, given the high uncertainty about the potential impact of climate change, would come with little cost penalty. This would not only allow a lower contribution of Europe to temperature increase but could also offset higher emissions in other world regions, compensating for inequitable historical emissions.

\newpage

\includepdf[pages=-]{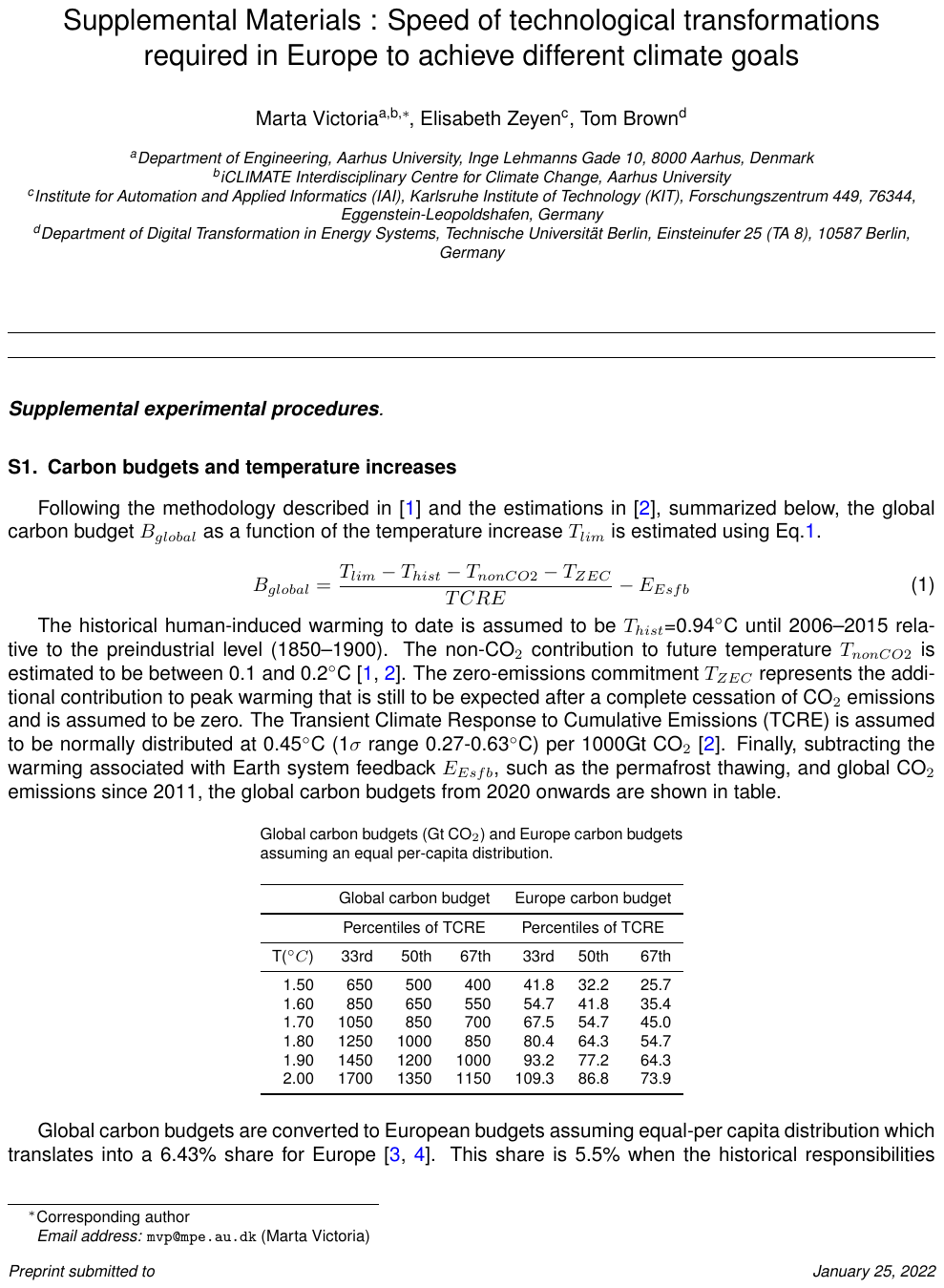}

\end{document}